\documentclass[a4paper,fleqn]{cas-sc}

\usepackage[authoryear,longnamesfirst]{natbib}

\def\tsc#1{\csdef{#1}{\textsc{\lowercase{#1}}\xspace}}
\tsc{WGM}
\tsc{QE}
\tsc{EP}
\tsc{PMS}
\tsc{BEC}
\tsc{DE}
\usepackage{lineno}
\usepackage[official]{eurosym}

\newcommand{\sMTOW}{\ensuremath{\sqrt{\mbox{MTOW}}}\xspace}
\newcommand{\TC}{\mathit{TC}}
\newcommand{\AFDR}{\mathit{AFDR}}
\newcommand{\AD}{\mathit{AD}}
\newcommand{\TR}{\mathit{TR}}
\newcommand{\ACD}{\mathit{ACD}}
\usepackage{array}

\usepackage[font=small,labelfont=bf,labelsep=period,
justification=centerlast]{caption}

\usepackage[caption=false,font=footnotesize]{subfig}

\usepackage{floatrow}
\usepackage{soul}  
\usepackage{multirow}

\usepackage{xspace}

\usepackage{units}

\begin{document}
\let\WriteBookmarks\relax
\def\floatpagepagefraction{1}
\def\textpagefraction{.001}
\shorttitle{Economic severity of Air Traffic Flow Management regulations}
\shortauthors{Delgado et~al.}

\title [mode = title]{Estimating economic severity of Air Traffic Flow Management regulations}                      



\author[1]{Luis Delgado}[orcid=0000-0003-4613-4277]



\address[1]{University of Westminster, Architecture and Cities, 35 Marylebone Road, London, GB, NW1 5LS}

\author[1]{G\'erald Gurtner}[orcid=0000-0003-2006-4653]

\author[2,1]{Tatjana Boli\'c}[type=editor,
    orcid=0000-0002-5530-7080]
\cormark[1]


\address[2]{Universit\`{a} degli Studi di Trieste, Dipartimento di Ingegneria e Architettura, Via A. Valerio 10, 34127 Trieste, Italy}

\author[2]{Lorenzo Castelli}[orcid=0000-0002-5621-0934]


\cortext[cor1]{Corresponding author, t.bolic@westminster.ac.uk, work performed while at Universit\`{a} degli Studi di Trieste \\Declarations
of interest: none.}


\begin{abstract}
The development of trajectory-based operations and the rolling network operations plan in European air traffic management network implies a move towards more collaborative, strategic flight planning. This opens up the possibility for inclusion of additional information in the collaborative decision-making process. With that in mind, we define the indicator for the \emph{economic risk} of network elements (e.g., sectors or airports) as the expected costs that the elements impose on airspace users due to Air Traffic Flow Management (ATFM) regulations. The definition of the indicator is based on the analysis of historical ATFM regulations data, that provides an indication of the \emph{risk of accruing delay}. This risk of delay is translated into a monetary risk for the airspace users, creating the new metric of the \emph{economic risk} of a given airspace element. We then use some machine learning techniques to find the parameters leading to this economic risk. The metric is accompanied by an indication of the accuracy of the delay cost prediction model. Lastly, the economic risk is transformed into a qualitative \emph{economic severity} classification. The economic risks and consequently economic severity can be estimated for different temporal horizons and time periods providing an indicator which can be used by Air Navigation Service Providers to identify areas which might need the implementation of strategic measures (e.g., resectorisation or capacity provision change), and by Airspace Users to consider operation of routes which use specific airspace regions.

\end{abstract}
%



\maketitle

\section{Introduction}\label{sec:intro}

After the reduction and stagnation in the period between 2008-2010, European air traffic started to increase again in 2011, bringing the same trend to the delays \citep{Eurocontrol2017}, an occurrence everyone prefers to avoid. At the time of writing, the aviation is experiencing an unprecedented decrease in traffic\footnote{Over 90\% decrease in traffic in April 2020, and still less than 50\% in August and September 2020 in Europe.} due to the COVID-19 pandemic, meaning that congestion and consequently delays are currently not a problem. Draft traffic scenarios\footnote{See https://www.eurocontrol.int/covid19\#traffic-scenarios} produced in April 2020 envisioned traffic increase throughout the remainder of the year, growing to the levels 30-20\% lower than in 2019. The September update to traffic scenarios, and the actual numbers show that the forecast was rather optimistic as the traffic is still at about 40\% of 2019 levels. The current scenario envisions very slow recovery in the early months of 2020 to about 50\% of the previous traffic. Hopefully, the vaccinations and other measures will enable slow return to normal which could lead to traffic growth later on. With traffic increase, the congestion and delays will return.

In the current European ATM, airlines enjoy high flexibility in the flight planning process - they are required to submit the flight plans from 120 to 3 hours before the flight, which gives them the opportunity to account for various uncertainty factors, like the weather, or aircraft frame availability~\citep{NM2018}. As a consequence, the accurate load on the airspace network (i.e., how many flights, where and at what time) is known on the day of operations. Conversely, the Air Navigation Service Providers (ANSPs) start preparing their capacity offer a year to six months before, considering historical traffic levels and staff availability. This mismatch in the planning horizons can result in capacity-demand imbalances, generating delay. Earlier sharing of information on flight intentions, or on the economic impact of flying through certain airspace at specified times could help both airlines and ANSPs obtain better demand-capacity management in the collaborative manner. 

As mentioned above, significant portion of delays is the consequence of the demand-capacity imbalance. When excess demand is expected, if capacity can not be adjusted, the ANSPs and Network Manager (NM) agree on ATFM measures that reduce the demand over the overloaded part of the network. A common ATFM measure is the issuing of regulations at specific traffic volumes (which can be usually identified as a portion of airspace), stating the start time, duration and agreed capacity. ATFM delays are imposed on flights which plan to use those resources, considering a First Planned - First Served principle as defined by the Computer Assisted Slot Allocation (CASA) algorithm \citep{CFMUmanual}. This delay is assigned to a flight, as ground delay prior to departure, ensuring that flows are smoothed over the congested area. ATFM delays can be caused by either airport or en-route related issues, which could be due to several different reasons. In the past five years, en-route delay accounted for 50–60\% of total ATFM delay, apart from 2018 and 2019 when en-route contributed to 75\% and 72\% of delay, respectively. The rest of ATFM delay was accrued at the airports \citep{Eurocontrol2020}. The major share of the en-route ATFM delay in the last years was due to the air traffic control (ATC) capacity, and ATC staffing, meaning that the available capacity of a portion of airspace was below the expected traffic demand.

The average ATFM delay between 2015 and 2017 remained stable at around 1.5 minutes per flight, and markedly increased in 2018 and 2019 to 2.33 and 2.18  minutes per flight respectively, which is very close to the maximum of 2.88 minutes recorded in 2010 (which had 13\% less traffic than 2018) \citep{Eurocontrol2019,Eurocontrol2020}. In 2019 there was an increment of 5\% of the number of flights issued ATFM delay with respect to 2018, with a total of over 1.4 million flights delayed, 40\% of those with more than 15 minutes of delay~\citep{Eurocontrol2020}. Considering an average ATFM delay cost of 100 \euro{}/min \citep{Cook2015}, total ATFM delay accrued in these years would imply delay costs to the airspace users that exceed 1 B\euro{} per year. As described in \citet{Cook2015}, there is, however, no direct linear relation between delay and cost of the delay. Therefore, detailed analysis needs to be carried out to gain understanding on the economic impact the regulations issued in different airspace regions have on flights~\citep{controller2015}.

Previous research has focused on the identification of regions with capacity-demand imbalance and measures to be applied pre-tactically, for example, managing the take-off times of flights to reduce congestion in the airspace \citep{NOSEDAL2014171,NOSEDAL201511}, adjusting dynamically the sectorisation to adjust capacity to demand \citep{TANG201289}, or balancing both  modification of trajectories to adjust demand, and opening schemes\footnote{Opening scheme determines which sectors are open, operational, at a certain time.} to adjust sectorisations to demand \citep{XU2020359}. Further work linked with pre-tactical decision making explored managing the ATFM delay distribution in order to reduce propagation of delay and improve airport slot adherence \citep{Ivanov17} and analysing trade-offs between efficiency and fairness in allocating ATFM delay when capacity at the destination airport is reduced, taking into account both flight and passenger indicators \citep{Delgado20}. Several papers address the strategic flight (and pre-tactical) planning striving to redistribute the traffic across the network to respect the airspace and airport capacities \citep{Bolic2017a}, or to achieve redistribution through peak-load pricing \citep{bolic2017peak}, or through modulation of charges \citep{Jovanovic2014}. The mentioned studies assume that sector and airport capacity values are fixed and not modifiable. This constraint is instead relaxed in the context of the “Coordinated capacity ordering and trajectory pricing for better-performing ATM” (COCTA) project where the balancing between air traffic demand and airspace capacity is achieved by applying economic instruments on the demand as well as on the capacity side \citep{ivanov2019,starita2020}. In this latter case, the authors mention the possibility that the NM can make capacity "orders". Based on scheduled traffic information and accounting for a portion of non-scheduled demand, NM can request one or more ANSPs to increase, or decrease the nominal capacity in some specific sectors to obtain better personnel management and therefore a reduction of ANS provision costs. 

Other research has focused on understanding the impact different factors have on delay and congestion \citep{aerospace5040109}. However, further efforts are required to enable the consideration of congestion and its impact strategically. The development of trajectory based operations and the rolling network operations plan provides this further collaborative environment which could be used as part of the strategic flight planning, opening the possibility for inclusion of additional information in the collaborative decision-making process~\citep{MasterPlan2020}. \citet{Xu2020} propose a collaborative ATFM framework for pre-tactical time horizon, where the airlines can decide how to deal with the saturated airspace elements, based on their own cost calculations. 

Having \emph{strategic} collaborative flight planning in mind, in this paper, we define  quantitative (economic risk) and qualitative (economic severity) indicators to assess the expected impact of the costs ATFM regulations would impose on airspace users. In particular, we present a methodological framework to estimate these metrics and we apply it to real data. As a part of this framework, a model to estimate average cost of ATFM delay as a function of average assigned delay is developed and presented. This historical analysis of the economic severity of ATFM regulations provides an initial strategic indication, which can be used by ANSPs to identify regions more prone to having a higher economic impact in the system. This might trigger the development of mitigation strategies such as new sectorisations and/or procedures. Further, a more detailed view can be used by airspace users as an indication of regions prone to generate a ``costly'' disruption of their operations and, once again, consider strategic mitigation actions such as planning flights to avoid inconvenient regions (e.g., re-routing, flight level capping) or, if possible, modifying planned buffers to reduce this expected impact.

The work presented in this paper is part of the ADAPT project which proposes a solution to enhance predictability while at the same time preserving and quantifying flexibility to be used strategically. Estimating the economic severity of ATFM regulations improves the understanding of the economic effect of hotspots (i.e., saturated portions of network) on airline operations.

Section~\ref{sec:methodology} presents the methodological framework used to devise the economic severity of airspace sectors. Section~\ref{sec:computation} presents the computations performed on historical data in order to create the economic risk and severity indicator. It comprises the data sources (Section~\ref{subsec:data}), the estimation of cost of delay  as a function of the average delay experienced by flights at given sectors (Section~\ref{subsec:avgcostestimation}), and the computation of the economic severity of the airspace (Section~\ref{sec:severity}). The conclusions are presented in Section~\ref{sec:conclusions}.


\section{Methodology\\ Framework for the definition of the economic severity of sectors}\label{sec:methodology}

The methodology section describes the framework to estimate an economic severity of different sectors/airspace which can be used in strategic, collaborative flight planning. The detailed description of data needed and specifics of calculations are then given in the Section \ref{sec:computation}. The economic severity can be calculated for different time-frames, which is also discussed later on. The economic severity framework is composed of following steps:

\begin{enumerate}
    \item An analysis of the impact of regulations by estimating the total expected cost generated by a regulation at a given sector ($\TC(s)$). This process is described in Section~\ref{subsec:tc}. The total expected cost depends on:
    \begin{itemize}
        \item The number of flights expected to be affected by a regulation issued on a sector (delayed flights),
        \item The probability of affected flights being assigned delay, and the magnitude of delays (based on average delay issued for each regulation), and
        \item A transformation of the expected total delay into cost, to move from classical flight-delay metrics to cost-related indicator.
    \end{itemize}
    
    \item An estimation of the total economic risk of a sector ($\TR(s)$), a quantitative measure, by considering the regulations issued at sectors and their impact (see Section~\ref{subsec:er}). The total economic risk of a sector depends on:
    \begin{itemize}
        \item The expected cost of a regulation at a given sector ($\TC(s)$), 
        \item The probability of a regulation being issued at a given sector while it is in operation, and
        \item The probability of a sector being in operation.
    \end{itemize}
    
    \item Transformation of the total economic risk of a sector into a qualitative, economic severity indicator, the use of which can be easier by decision makers. This is detailed below, in Section~\ref{subsec:severity}.
\end{enumerate}

\subsection{Total expected cost of a regulation at a sector ($\TC(s)$)} \label{subsec:tc}

As described in \cite{Cook2015}, the relationship between delay and cost is not linear, and depends on several factors, such as aircraft size or location where delay is accrued. Section~\ref{subsec:avgcostestimation} presents more details on this relationship, and describes the estimation of average cost of delay for a given regulation. 

As the flight execution time horizon approaches, the amount of available information increases, allowing more accurate predictions of the expected cost of a regulation to be incurred by airspace users (e.g., better estimations on the distribution of the issued delay, or even of the expected average delay could be obtained). However, in this paper, we focus on the \emph{strategic phase} and hence, we consider only information coming from historical data (usually, the only data available strategically). Historical data could be at a high aggregation level and some parameters, such as the type of regulation might not even be available. 

With all these considerations, we define the total expected costs of a regulation at sector $s$ ($\TC(s)$) as:

\begin{equation}
\label{eq:1}
\TC(s) = \AFDR(s) \times \AD(s) \times \ACD(s)
\end{equation}
where:
\begin{itemize}
    \item $\AFDR(s)$ denotes the average number of flights which are assigned delay when a regulation is issued at sector $s$: When the flight plan crosses an active regulation, there is a probability that some delay will be assigned to it (e.g., around 63\% of flights which enter at least one regulation get delay assigned according to an analysis of historical data (three months of data from AIRAC\footnote{Aeronautical information regulation and control (AIRAC) period, always composed of 28 days. AIRAC 1409 stands for 28 days, mostly falling in September(09) of 2014 (14).} 1409,1702,1709)).
    
    \item $\AD(s)$ is the average delay per delayed flight: The total amount of delay that is assigned by a given regulation varies as a function of different parameters which are related to the regulation reason and its duration, location and severity. From historical data, it is possible to estimate the individual delay that a given flight might experience and then transform this estimation into expected cost considering the flight characteristics. However, this information is usually available only on the day of operations. Therefore, at strategic level, instead of focusing on individual flights, we consider the average delay that each historic regulation assigned.
    
    \item $\ACD(s)$ is the average cost per minute of delay: Once the expected delay that a regulation generates is estimated, we need to assess the economic cost that airlines will incur due to this delay. As described in \citet{Cook2015} and \citet{controller2015}, this relationship is not linear and depends on several factors, such as aircraft size or location where delay is accrued. Section~\ref{subsec:avgcostestimation} describes in detail the transformation of average assigned delay to average assigned cost.
\end{itemize}

\subsection{Total economic risk of a sector ($\TR(s)$)} \label{subsec:er}

After estimating the total expected cost of a given regulation in a given sector ($\TC(s)$), we turn to estimating the associated risk ($\TR(s)$) of experiencing this cost.  $\TR(s)$ depends on the probability of sector $s$ being regulated ($P_s$), leading to:

\begin{equation}
\label{eq:2}
\TR(s) = \TC(s) \times P_s(\mbox{regulated})
\end{equation}
The total risk for each sector can be computed for different time periods (e.g., yearly, quarterly or monthly) and represents a quantitative measure of risk. Note that the adequate time-horizon could vary as a function of the intended use of data by different stakeholders\footnote{The stakeholders in this case are airlines and ANSPs.}.

Airspace sectors are not always active. The configuration of the airspace varies according to the opening scheme which adapts the provided capacity\footnote{Through the choice of number of sectors to open and their general characteristics.} to the expected demand considering operational constraints. Therefore, the probability of element $s$ being regulated can be defined as:

\begin{equation}
\label{eq:3}
P_s(\mbox{regulated}) = P_s(\mbox{regulated}|\mbox{open}) \times P_s(\mbox{open}).
\end{equation}

The probability of a sector being `open' refers to the fraction of time sector $s$ is active  $\Delta T_o$ (i.e., open), within the time horizon $T_H$ of the analysis. 

\begin{equation}
\label{eq:3_1}
P_s(\mbox{open})=\frac{\Delta T_o}{T_H},
\end{equation}

In order to calculate the probability of a sector being regulated, we merge the data from two data sources: one containing information on the regulations (their location and duration), and another containing the information on sector opening times (from historical opening schemes). We estimate the probability of a sector $s$ being under regulation if it is open, by computing:

\begin{equation}
\label{eq:4}
P_s(\mbox{regulated}|\mbox{open}) = \frac{\Delta T_r}{T},
\end{equation}
where $\Delta T_r$ is the duration of the regulation and $T$ the time during which the sector is open\footnote{In case $\Delta T_r$ is greater than $T$, we limit $\Delta T_r$ to be equal to $T$.}.

Therefore, combining Equations \ref{eq:1}, \ref{eq:2}, \ref{eq:3} and \ref{eq:4} we obtain that the economic risk associated with a sector $s$ in a given time-frame can be estimated as:

\begin{equation}
\label{eq:all}
\TR(s) = \AFDR(s) \times \AD(s) \times \ACD(s) \times \frac{\Delta T_r}{T} \times P_s(\mbox{open})
\end{equation}

As presented, the proposed quantitative metric of economic risk only requires high-level aggregated data: the average number of flights with delay issued due to regulations in the sector, the average delay per delayed flight and the average cost per minute of delay, to estimate the cost of the regulations in the sector; and the ratio of time the sector is regulated over the time it is open, times the probability of being in operation.

\subsection{Economic severity of a sector} \label{subsec:severity}

The total economic risk provides a quantitative value for each sector, but we consider that the severity should be described in a qualitative manner. Qualitative description eases understanding of criticality of sectors in a more operational approach. Grouping numerical values in a small number of categories reduces complexity, especially if groupings are created with a specific usage in mind. Each category can then be given a qualitative description which can help the stakeholders in their operational decision making. 

Different categorisations could be devised based on the total risk values. Of course, for the proposed metric to be put in use, the input from stakeholders is required to capture their preferences. In this paper, we are proposing and testing the metric, and as an example, we compute the economic severity of a sector by ranking the sectors considering their economic risks and classifying them in one of five severity category based on their economic risk  percentile as: very low, low, medium, high and very high severity.

Table~\ref{tab:economic_severity} presents the relationship between the percentile of the total risk and the severity that we have used as an example in this paper. For this example, we use categories assigned across quantiles that are not uniformly distributed. We chose this crisp set of categories for simplicity. 
Note that in this particular severity categorisation, severity is defined relative to the total risk of the set of sectors included in the analysis and being classified. Therefore, for example, the sectors with the higher total risk (i.e., the top 10\% from the set being classified) will be assigned a `very high' category regardless of the absolute value of their risk. This type of categorisation needs to be taken into account if we want to compare economic severities which have been assigned from different datasets. A definition which considers the absolute risk value could also be designed, but in this case, input from stakeholders is required in order to consider their needs and preferences.

\begin{table}[ht]
    \centering
    \caption{RELATIONSHIP BETWEEN PERCENTILE OF TOTAL ECONOMIC RISK OF SECTORS AND ECONOMIC SEVERITY USED AS AN EXAMPLE}
    \begin{tabular}{c|c}
    Percentile of total economic risk & Economic severity  \\
    \hline
    [0.0 - 0.1] & Very Low \\
    (0.1 - 0.3]	& Low \\
    (0.3 - 0.6]	& Medium \\
    (0.6 - 0.9]	& High \\
    (0.9 - 1.0] &	Very High \\
    \end{tabular}
    \label{tab:economic_severity}
\end{table}

\section{Economic risk and severity computation}
\label{sec:computation}

In this section we describe in detail used data sources and demonstrate the application of the presented framework on a specific dataset. The main contribution here is the use of machine learning techniques to  transform the delay information into costs. The model appropriateness, statistical significance and errors the chosen model brings are very important for understanding of information the proposed indicators (economic risk and economic severity) convey to users.

\subsection{Data sources}\label{subsec:data}

The study presented in this paper uses two datasets - \textbf{\emph{aggregate regulations and flight delay}}. The \textbf{\emph{aggregate regulations}} is the main, historical data set used for calculation of probabilities and delays, as  explained in the Section~\ref{sec:methodology}. Conversely, the \textbf{\emph{flight delay}} dataset enables us to  estimate the relationship between average delay and average cost. 

\begin{figure}[h]
    \centering
    \includegraphics[width=0.4\textwidth]{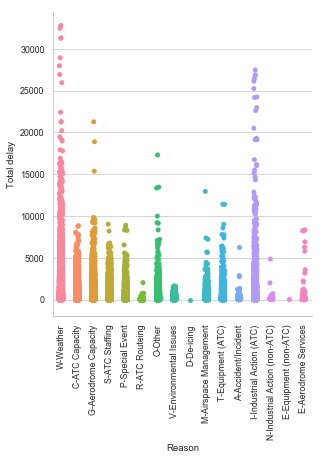}
    \caption{Total delay assigned per regulation per ATFM reason.}
    \label{fig:tdelay_reg_reason}
\end{figure}

Let us begin with the \emph{aggregate regulations} dataset that contains historical data on the ATFM regulations and their impact, along with the airspace usage (i.e., configurations in operation). This dataset is used to compute the economic risk and severity of different airspace regions. The regulation data is sourced from the \emph{Daily ATFCM Summary} files, from the EUROCONTROL’s Network Manager ATFCM statistics website, and the airspace configuration data and the sector opening times from the \emph{Demand Data Repository 2 (DDR2) environmental files}. The data covers three years: 2016, 2017, and 2018. The \emph{ATFCM daily summary} reports contain ten different reports on the impact of regulations on different parts of the network and stakeholders, in varying levels of aggregation. For our purposes, we use data from three reports, more specifically the following fields: 
\begin{itemize}
    \item Date of regulation, 
    \item Regulation name, 
    \item Number of regulated flights (those affected by the regulation),
    \item Number of delayed flights (those assigned delay due to the regulation), 
    \item Total delay (minutes), 
    \item Average delay per regulated flight (minute/flight), 
    \item Average delay per delayed flight (minute/flight), 
    \item Reference location (unique name of the sector, navigation point or airport over which the regulation is imposed),
    \item Location type (airport or en-route),
    \item Regulation duration (in minutes, can be negative in case the regulation has been cancelled before its scheduled start time), and 
    \item Regulation reason\footnote{We use terms "regulation reason" or "regulation type" interchangeably.} description.
\end{itemize}
There are 16 regulation reasons, some can be applied only at airport, or en-route, and some can be applied in both categories. Regulations issued due to different reasons tend to assign different total amount of delay, as shown in Figure~\ref{fig:tdelay_reg_reason}. This is related to the link between the regulation type and other parameters such as duration of the regulation or capacity reduction.

The \emph{DDR2 environmental data} contains two file types that report the available configurations and the opening times of the configurations in each AIRAC cycle. From these two files, the configurations can be extracted, and consequently obtain the information on openings and duration of openings of individual sectors. This data is merged with the regulation data to form our \emph{aggregate regulation} dataset, containing three years of data. 

\begin{figure*}[!t]
\centerline{
\subfloat[Delay]{\includegraphics[width=0.45\linewidth]{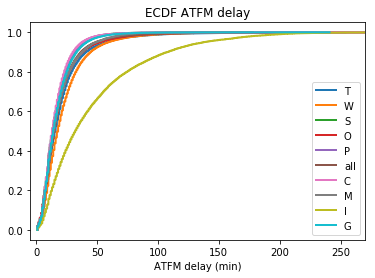}%
\label{subfig:ecdf_atfm_delay}}
\hfil
\subfloat[Cost of delay]{\includegraphics[width=0.45\linewidth]{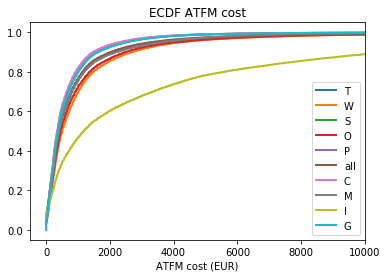}%
\label{subfig:ecdf_atfm_cost}}}
\caption{Experimental cumulative distribution (ECDF) of delay and cost of delay as a function of ATFM regulation reason.}
\label{fig:ecdf_atfm}
\end{figure*}

Finally, the presented framework requires the estimation of average cost of the ATFM delay experienced by airspace users, the $\ACD(s)$ mentioned in Section~\ref{subsec:er}. This is a complex relationship and a more detailed flight-by-flight dataset is needed to estimate the individual cost of delay of flights. Thus, the second dataset (\emph{flight delay}) contains data on individual regulated flights, aircraft type used, and assigned ATFM delay, as well as the name of regulation that caused the delay. These have been sourced from the \emph{DDR2 flight data}. These detailed data were available only for the  AIRACs 1313-1413 and 1702, 1709. Initial analysis of this data is presented in Figure~\ref{fig:ecdf_atfm}, depicting how the expected delay (and associated cost) for a given regulated flight might differ across regulation type. These relations are explored in detail in the next section.   

\subsection{Estimation of cost of average ATFM delay} \label{subsec:avgcostestimation}

As presented in Equation~\ref{eq:all}, the economic risk of a sector is computed based on the expected total cost generated by ATFM regulations issued in that sector. In order to obtain the expected total costs, we need to model the relationship between ATFM delay and the cost experienced by airspace users. The cost of delay for a given flight is very well documented in \citet{Cook2015} and depends on several factors, in particular:
\begin{itemize}
    \item The phase of the flight where delay is produced: at-gate, taxi, or airborne,
    \item The type of aircraft: which affects operational parameters, such as fuel consumption, number and type of crew, or number of passengers, among others,
    \item The type of airline: \citet{Cook2015} provides low, baseline and high estimated cost profiles (that can be assigned to different types of airlines),
    \item Passengers on-board and type of operations: number of passengers and if they are point-to-point or connecting passengers.
\end{itemize}

The different costs that an airspace user experiences due to delay can be divided in:
\begin{itemize}
    \item Passenger costs, which are further decomposed in hard costs (due to duty of care, compensation due to Regulation 261 \citep{EU2004} and missed connections handling) and, soft costs (due to the potential loss of market share due to delay experienced by passengers),
    \item Non-passenger costs, composed of: crew, maintenance and fuel costs,
    \item Reactionary costs due to the propagation of delay.
\end{itemize}

In this research, we consider the cost of delay as the ``at-gate'' costs (as the ATFM delay is performed at gate before push-back), using the baseline\footnote{Baseline is a mid-point cost scenario presented in \cite{Cook2015}.} cost profile which includes all the above-mentioned estimated costs. As the intended use of the metric is in the strategic collaborative flight planning, we use baseline cost profile.

\begin{figure}[h]
    \centering
    \includegraphics[width=0.5\linewidth]{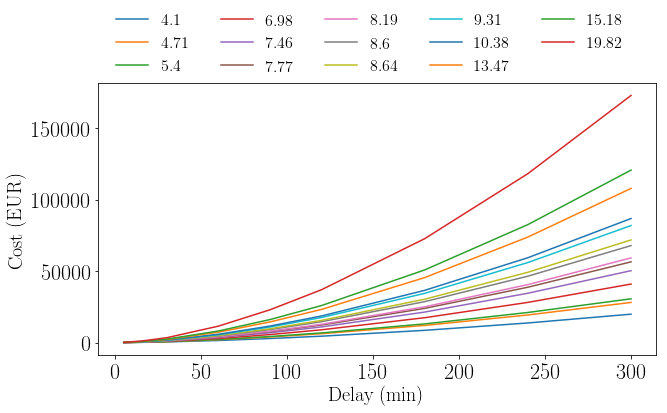}
    \caption{Cost of delay at gate for different aircraft types (presented as different \sMTOW), data from AIRACs 1313-1413, 1702, 1709.}
    \label{fig:cost_mtow}
\end{figure}

The relationship between delay and cost is non-linear and it has been demonstrated that there is a good correlation between the square root of maximum take-off weight (MTOW) of an aircraft and the cost experienced over the amount of delay~\citep{Cook2015}. This is due to the link between \sMTOW and the parameters which affect this cost of delay (e.g., number of passengers or fuel consumption). Hence, if no details are available on the flight (e.g., number of passengers on board) and type of airline operations, the MTOW of the aircraft can be used to estimate the cost of delay. Figure~\ref{fig:cost_mtow} shows the costs of delay per delay duration for different aircraft types (presented as the \sMTOW) computed following~\cite{Cook2015}. 

From the presented cost of delay relationships, one can observe that the average delay in a single flight delay-cost relationship cannot be used as there are non-linearities:
\begin{itemize}
    \item Early flight (negative delay) does not yield, in general, any advantage for the airline, and
    \item The cost grows non-linearly with the delay, i.e., the cost of higher delays is proportionally much higher than the cost of small delays.
\end{itemize}

These non-linearities can be captured by the following equation:

\begin{equation} \label{eq:quad}
c_d(\delta)=
\begin{cases}
\alpha \delta + \beta \delta^2\quad, \mbox{if}\  \delta>0\\
0 \quad, \ \mbox{otherwise}
\end{cases}
\end{equation}
where $c_d$ if the cost of delay given the delay $\delta$, and $\alpha$ and $\beta$ are two positive parameters.

Due to the non-linearity and the dependence on operational parameters such as \sMTOW, in general, the full distribution of delay and aircraft experiencing the delay is needed in order to properly estimate the cost of delay. However, the full distribution of delay per aircraft type is not always available, especially at the strategic phase. For this reason, the transformation of delay into cost will be estimated considering the average cost of delay ($\ACD(s)$), which is available from historical datasets. Therefore, we need to provide a model able to infer the average cost of delay using degraded information, i.e., only the average delay, and maybe the type (reason) of regulation, as these might be the only information available (or that can be estimated) strategically. We also want to have an estimation of the error incurred by considering only these predictors instead of using a more detailed dataset with individual delays and aircraft types.

We estimate the model presented in Equation \ref{eq:quad} using data on individual regulated flights contained in AIRACs 1313-1413 and 1702, 1709 (i.e., about 17 months). Dataset comprises a total of 1\ 029\ 699 flights regulated in 25\ 728 regulations (9\ 631 (37\%) due to ATC Capacity (C), 4\ 532 (18\%) due to Weather (W), 3\ 584 (14\%) due to Aerodrome Capacity (G), 1\ 912 (7.4\%) due to ATC Staffing (S), 7\ 981 (24\%) for all other reasons). For each flight we have the exact amount of ATFM delay that was assigned, the aircraft type, and the regulation.

For each of those regulated flights, the expected cost of delay has been computed considering their \sMTOW and the ATFM delay assigned following \cite{Cook2015}. In the next step, the delays and costs of individual regulated flights are allocated to the regulations based on the most penalising regulation principle, i.e., the regulation indicated as the most penalising regulation for each flight is considered to have that delay (and cost) allocated to it. Note that in some cases, regulations might not generate delay as they might not be the most penalising ones for any flight, if flights cross more than one regulation.

\begin{figure}[h]
    \centering
    \includegraphics[width=0.5\linewidth]{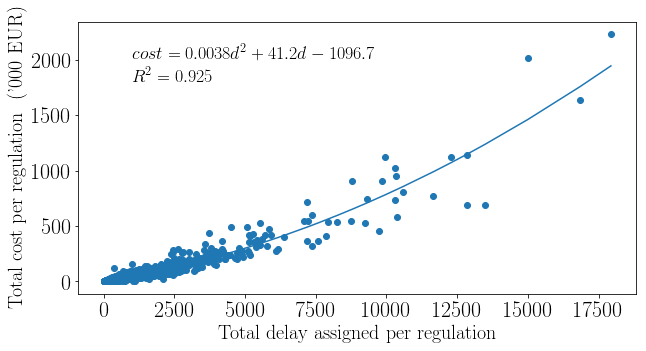}
    \caption{Total cost generated per regulation as a function of total delay assigned per regulation (AIRACs 1313-1413, 1702 and 1709).}
    \label{fig:cost_delay_fit}
\end{figure}

As expected, there is a relationship between the total delay that a regulation assigns and the cost that it creates for the airspace users. Figure~\ref{fig:cost_delay_fit} presents the total cost of the delay generated for each regulation as a function of the total delay assigned per regulation and how it can be approximated with a quadratic fit having an R$^2$ of 0.925.

In this research, the focus is on the estimation of the cost, based on the average delayed flight in a regulation, at the strategic level. As already mentioned, the characteristics of the traffic are not fully known strategically, and the regulation reason is usually not known strategically. Thus, we analyse the regulations by considering all the flights going through a given regulation having delay and averaging their delay and corresponding costs. However, as shown in Figure~\ref{fig:ecdf_atfm} the amount of expected delay can differ across different regulation reasons, due to the link between the regulation type, regulation duration and capacity reduction. Thus, even though strategically the regulation type is not known, the type of regulation is explored to understand if and how they increase the predictive level of the model.

\begin{figure}[h]
    \centering
    \includegraphics[width=0.5\linewidth]{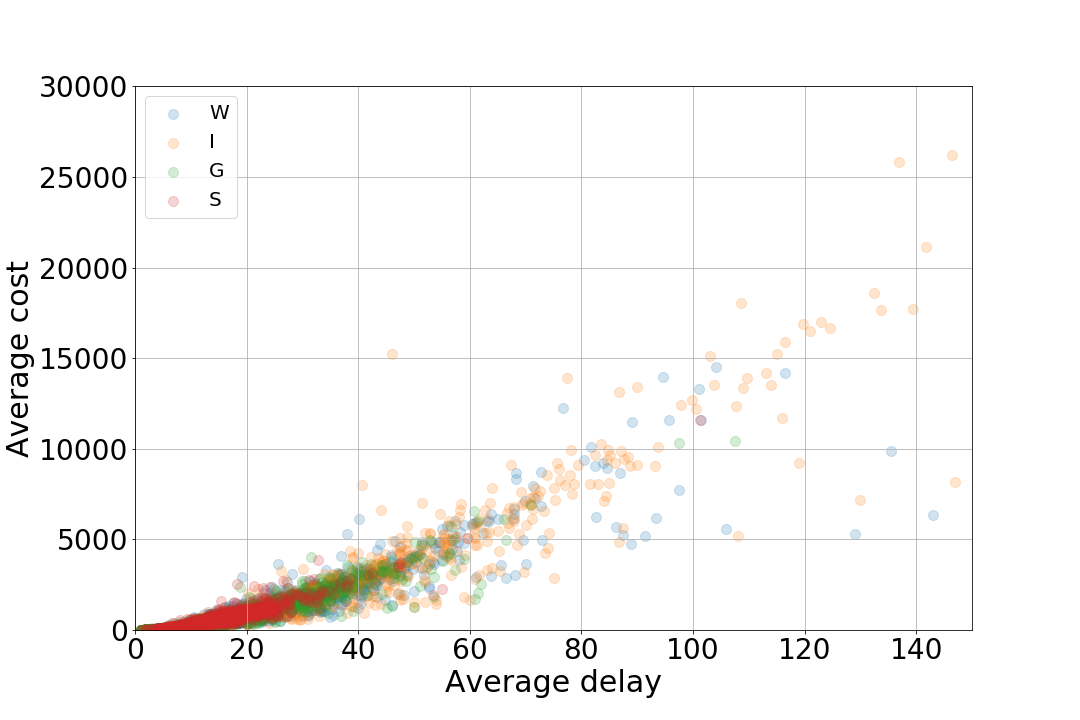}
    \caption{Average cost incurred by a flight with ATFM delay issued by a regulation versus the average delay issued by those regulations.}
    \label{fig:cost_reg_reason_plot}
\end{figure}

To explore the data, and in particular the impact of the type of regulations, we start by plotting the two variables: delay and cost restricted to the several regulation reasons (types). Figure~\ref{fig:cost_reg_reason_plot} shows the average delay and cost for four types of regulations (Weather (W), Industrial action (I), ATC staffing (S), and Aerodrome capacity (G) as those account for majority of total delay), with a few outliers removed. From the figure, it seems that the data exhibits the quadratic behaviour hypothesised in the cost function (see Equation~\ref{eq:quad}). Clearly, some costs are below the average behaviour at high delay. Moreover, it seems all the points of the different types of regulations are roughly aligned on the same curve, even if some of them explore the curve to a different extent (for instance industrial delays are more likely to be high, but so is their associated cost).

As shown in Figure~\ref{fig:ecdf_atfm}, the reason of a regulation might impact the probability of having a given amount of delay assigned, but here we are interested in the relationship between delay and cost. Thus, from a statistical point of view, the first question is whether or not we need the regulation reason variable to predict the cost, given that we have access to the average delay. To test this, we performed an ordinary least-square (OLS) regression, using as predictors the average delay and all dummy variables related to the different regulation types present in the dataset:

$$
c= \alpha \delta + \beta D_I + \gamma D_W + ...
$$

Given that there are a few outliers, we restricted the data to the points with a z-score \footnote{Z-score, or standard score, is a measure of how far a given data point is from the empirical mean in the dataset (measured in units of standard deviation). Very high z-scores indicate outliers.} smaller than 3 in cost and 5 in average delay. In all the following analyses, we control overfitting by using a technique common in machine learning and divide our dataset into two parts: the training part (in-sample), composed of 75\% of the dataset randomly selected, and the testing set (out-sample), composed of the rest. The relative fitness scores of the regressions in these two samples allow to control overfitting.

The dummy variables do not have a strong predictive power (see Table~\ref{tab:OLS_coef}). Only the reasons I, M, and O are found statistically significant, and it is likely more related to the magnitude of the delays of these regulations rather than to the structural difference in the cost-delay relationship. We are not interested in the former, since we are computing independently the average delay conditioned on the reason. The regulation reason might play a role in the estimation of the expected average delay (as presented in the previous section).

\begin{table}[ht]
    \centering
    \caption{COEFFICIENTS GIVEN BY THE OLS PROCEDURE WITH ASSOCIATED STANDARD DEVIATIONS (* denotes statistically significant coefficient at a 1\% level)}
    \begin{tabular}{c|c c}
    & Coefficient & Standard error  \\
    \hline
    Intercept &	-0.093 & 0.055 \\
    \hline
    $\delta$*	& 0.90	& 0.003\\
    \hline
    $D_C$	& 0.14	& 0.055\\
    \hline
    $D_G$	& 0.033	& 0.056\\
    \hline
    $D_I$*	& 0.16	& 0.059\\
    \hline
    $D_M$*	& 0.20	& 0.060\\
    \hline
    $D_O$*	& 0.22	& 0.057\\
    \hline
    $D_P$	& 0.10	& 0.057\\
    \hline
    $D_R$	& 0.069	& 0.060\\
    \hline
    $D_S$	& 0.12	& 0.056\\
    \hline
    $D_T$	& 0.080	& 0.059\\
    \hline
    $D_V$	& 0.0005 & 0.065\\
    \hline
    $D_W$	&-0.019	& 0.056\\
    \end{tabular}
    \label{tab:OLS_coef}
\end{table}

Given this result, we then consider a simple quadratic fitting between the average cost and the average delay, using all the data available, not restricted to any regulation type:

\begin{equation}
    \bar{c} = \alpha \bar{\delta} + \beta \bar{\delta}^2
\end{equation}

\begin{figure}
    \centering
    \includegraphics[width=0.5\linewidth]{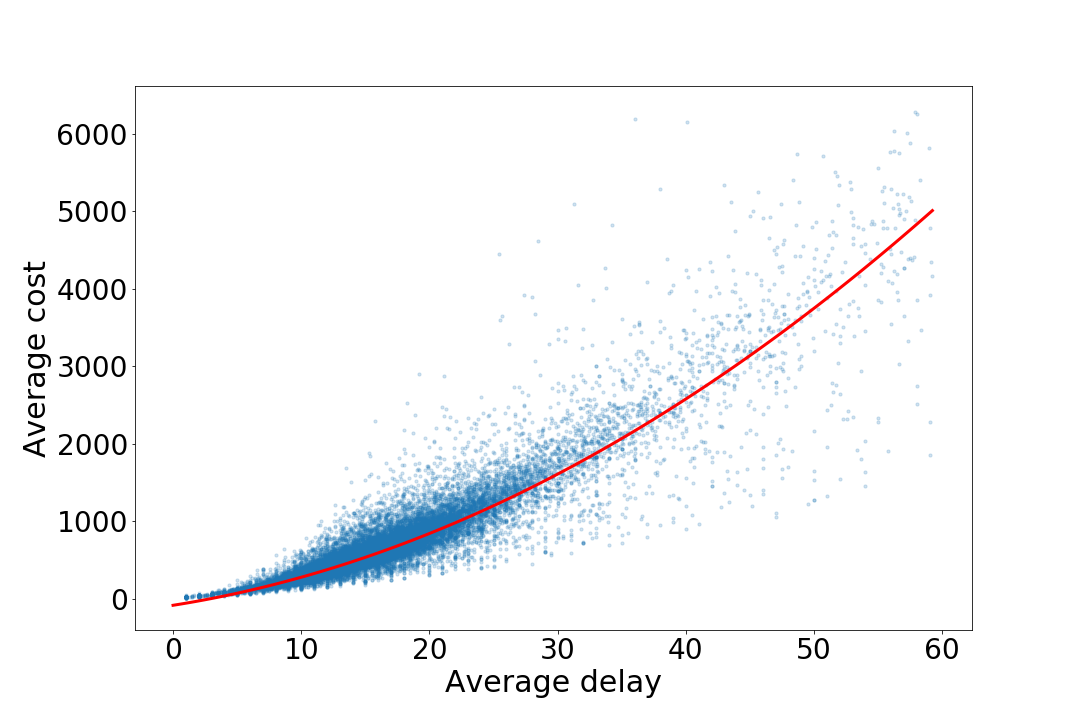}
    \caption{Quadratic fit with in-sample $R^2=0.93$ and out-sample $R^2=0.93$.}
    \label{fig:quadratic_fit}
\end{figure}

The regression fit is good, as shown in Figure~\ref{fig:quadratic_fit}, representing quite well the entire dataset, independently of the regulation type ($\alpha=23.1$ \EUR{} per minute, $\beta=1.05$ \EUR{} per minute square). Both in-sample and out-sample have $R^2$ of 0.93.

\begin{figure}[h]
    \centering
    \includegraphics[width=0.65\linewidth]{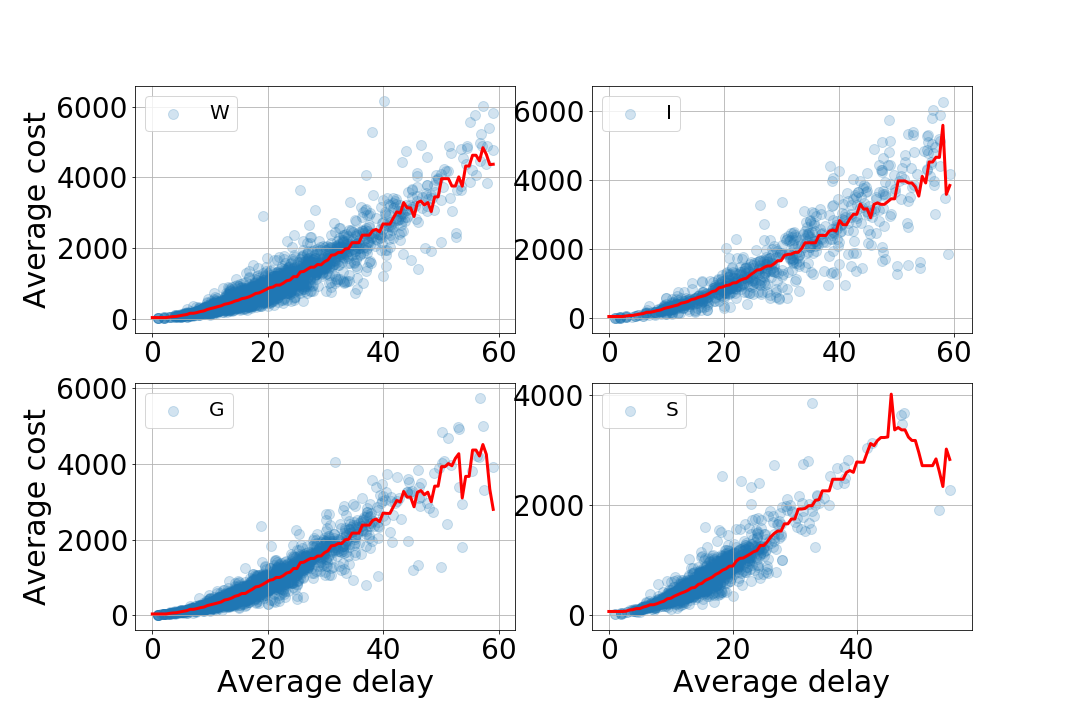}
    \caption{Average cost vs. average delay for different regulations. Red line is the result of the GBT fitting.}
    \label{fig:gbt_fitting}
\end{figure}

Even though the regression fit of the quadratic model is good, we wanted to further explore the exact impact of the regulation reason, in case we can find a better fitting model. To do so, a non-linear regression method, the gradient boosting tree (GBT), has been used. This technique is based on decision trees and is very powerful with categorical variables. When mixed with continuous variables, it allows to capture simultaneously different regression functions \citep{friedman_greedy}. The downside of decision trees is that they are prone to overfitting. The results of the GBT are illustrated in Figure~\ref{fig:gbt_fitting}, where we show the application of the tree, trained on all the (training) data, to different sub-samples corresponding to four regulation types (the same ones displayed in Figure~\ref{fig:cost_reg_reason_plot}). The figure shows that the GBT can capture different regressions for different regulations. However, the main differences lie in the high delay parts, which are overfitted.

\begin{figure}
    \centering
    \includegraphics[width=0.5\linewidth]{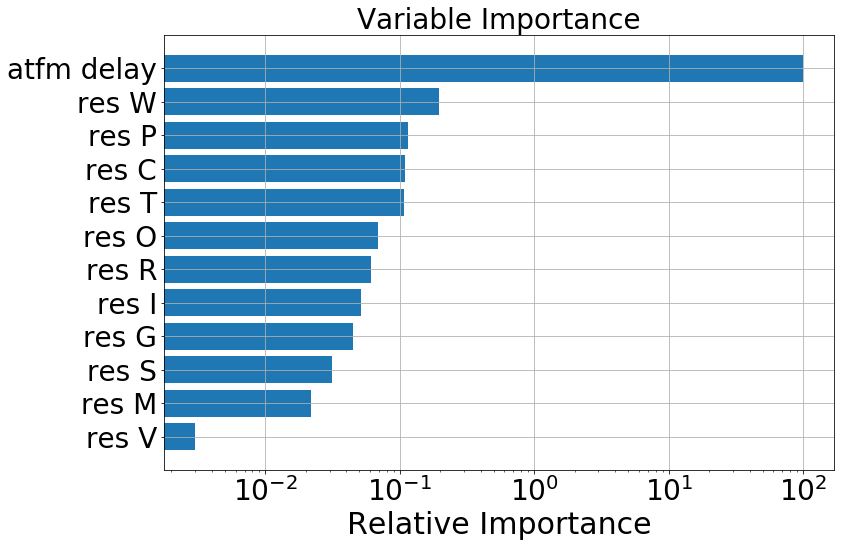}
    \caption{Relative importance of the predictors in the prediction of the average cost. The abscissa has a log scale, showing 0.1\% or less importance of regulation reason in the prediction.}
    \label{fig:relative_importance}
\end{figure}

To see the impact of the regulation on the tree, one can display the relative importance of the predictors within the trees inferred by the model, as shown in Figure~\ref{fig:relative_importance}. The figure has a logarithmic scale (in percentage), and it clearly shows that different reasons have a marginal impact in predicting the average cost, as opposed to the average delay itself. In conclusion, a model more complex than the quadratic regression based on the average delay is not justified, as the inclusion of information on the regulation type does not improve the predictive power of the model significantly.

\begin{figure}
    \centering
    \includegraphics[width=0.5\linewidth]{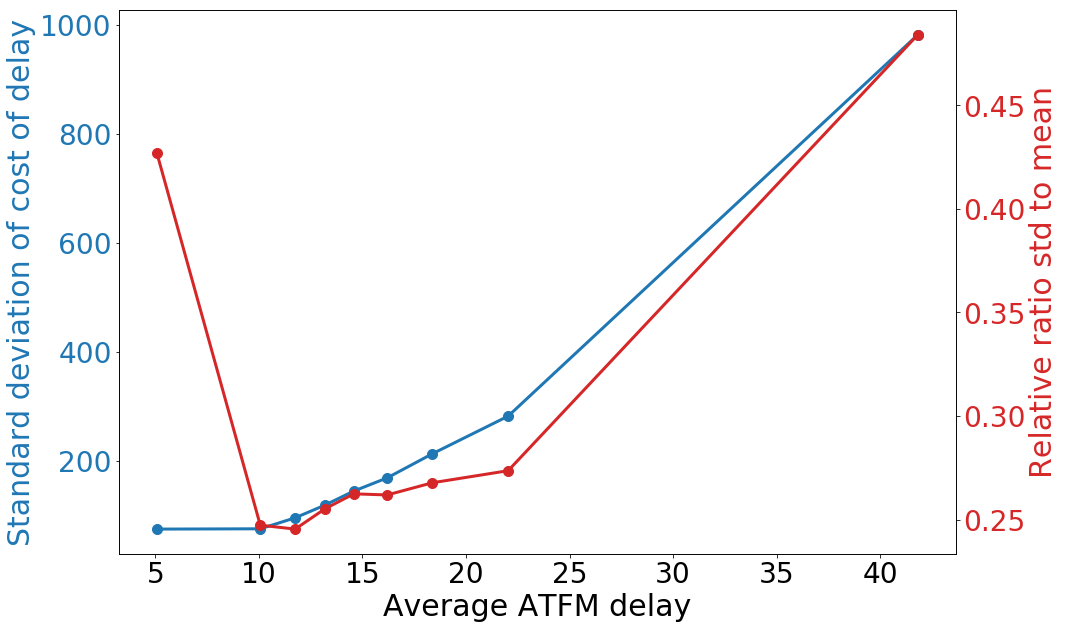}
    \caption{Standard deviation in the different quantile bins as a function of the average ATFM delay in the bin (blue), ratio between the standard deviation and the mean on the same bins (red).}
    \label{fig:std}
\end{figure}

However, since we also know that the average delay is not enough to represent all the variations of the cost, we compute a metric to indicate the potential error when using this approximated formula. For this, we use the standard deviation, as it is the most well-known and probably the most intuitive measure of dispersion for a distribution. We want an indication of the variance of the cost, based on a given average ATFM delay in an airspace, to estimate how typical is the average. To represent a typical standard deviation for this point, we decide to bin the data in equal parts using equally spaced quantiles. For each quantile, we compute the standard deviation of the cost of delay for the points within the quantile. This allows to have good statistics for each point, since they are computed on the same number of points. Figure~\ref{fig:std} shows the standard deviation computed on 10 quantiles and its evolution with respect to the centre of the bin. The deviation increases a lot when the cost of delay increases, from around 50\euro{} up to 1000\euro{}. Interestingly, it is also the case for the ratio between the standard deviation and the mean, which suggests the above model fails at high delays. It is particularly noticeable on the last point, where the error goes from less than 30\% (0.30 in Figure~\ref{fig:std}) to more than 45\%. Finally, it is interesting to note that the very low delays have a high relative error too. This is due to the fact that even when the average delay is close to 0, there is some dispersion in the data and thus the ratio of standard deviation over mean diverges. 

To sum up, based on the presented analyses, the quadratic model is chosen to estimate average costs from average delay. The model parameters and the model errors are given in the Table~\ref{tab:model}.

\begin{table}[ht]
    \centering
    \caption{AVERAGE COST AND STANDARD DEVIATION - MODEL ESTIMATES}
    \begin{tabular}{c|c c}
    &\multicolumn{2}{c}{Model} \\
    \hline
    \hline
    Average cost &	$\bar{c} = \alpha \bar{\delta} + \beta \bar{\delta}^2$ & $\alpha=23.1$ Euros per minute, $\beta=1.05$ Euros per minute square \\
    \hline
    \hline
    \multirow{ 10}{*}{Model error}	& Average ATFM delay (minutes)	& Std deviation cost (\euro{}), (percentage w.r.t mean)\\
    \cline{2-3}
    	& 1.0 -- 9.13	& 74.5 (43\%)\\
	\cline{2-3}
    	& 9.13 -- 11.0	& 75.1 (25\%)\\
	\cline{2-3}
    	& 11.0 -- 12.5	& 95.2 (25\%)\\
	\cline{2-3}
    	& 12.5 -- 13.9	& 118 (26\%)\\
	\cline{2-3}
    	& 13.9 -- 15.3	& 144 (26\%)\\
	\cline{2-3}
    	& 15.3 -- 17.1	& 169 (26\%)\\
	\cline{2-3}
    	& 17.1 -- 19.7	& 213 (27\%)\\
	\cline{2-3}
    	& 19.7 -- 24.4	& 292 (27\%)\\
	\cline{2-3}
    	& 24.4 -- 59.2	& 982 (48\%)\\
    \end{tabular}
    \label{tab:model}
\end{table}

\subsection{Economic risk and severity}\label{sec:severity}

\begin{figure*}[!h]
\centerline{
\subfloat[Average cost per delayed flight]{\includegraphics[width=0.45\linewidth]{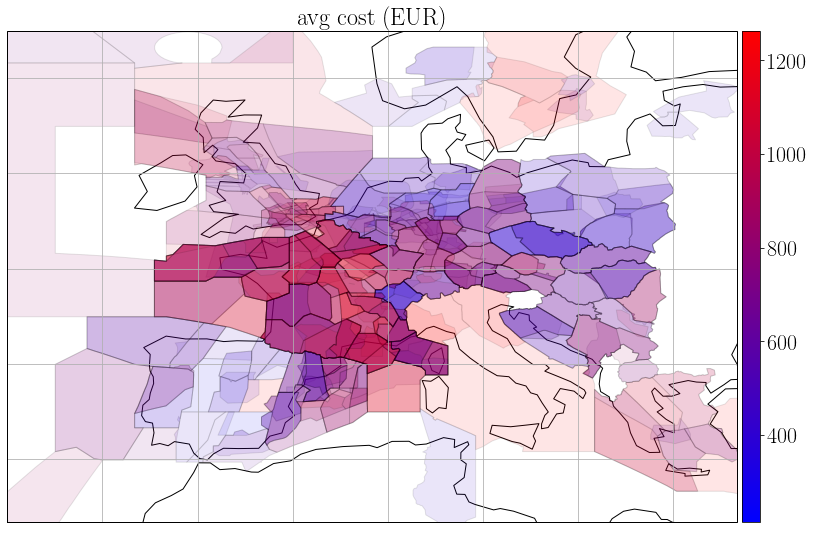}%
\label{subfig:avg_cost_per_sector}}
\hfil
\subfloat[Standard deviation of average cost]{\includegraphics[width=0.45\linewidth]{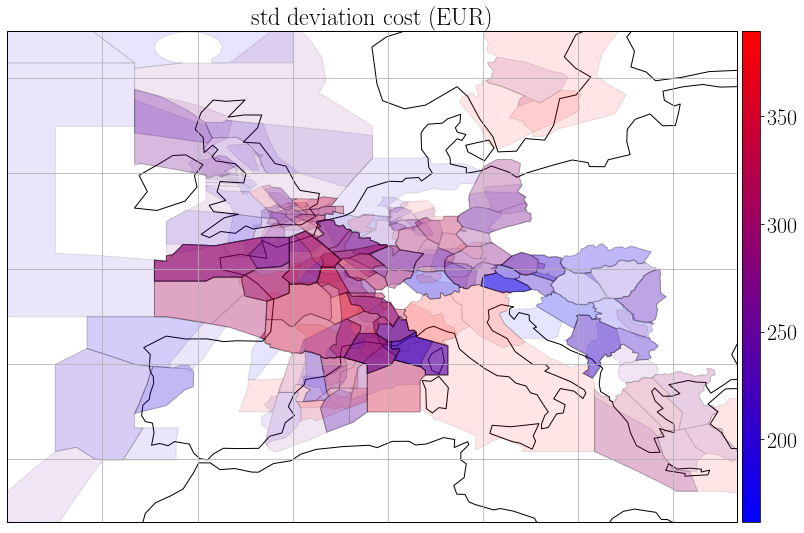}%
\label{subfig:std_cost_per_sector}}}
\caption{Estimated average cost per delayed flight with standard deviation of average cost per sector.}
\label{fig:avg_std_cost_per_sector}
\end{figure*}

Finally, for each historical regulation present in the aggregate regulation dataset, for which the average ATFM delay is available, the regression model to estimate the average cost and standard deviation of the cost have been applied. For each regulated sector we estimate expected average cost. We can then plot the sector map, colored according to the value of expected average cost, see Figure \ref{fig:avg_std_cost_per_sector}. 

Figure~\ref{subfig:avg_cost_per_sector}, shows the estimated average cost per delayed flight per sector. Note that some sectors might overlap in this representation. Some regions present higher average costs than others. This might be due to the type of airspace, traffic, or regulation types that were produced on those areas. Finally, in Figure~\ref{subfig:std_cost_per_sector} the standard deviation of the cost with respect to the estimates presented before are depicted. As discussed previously, there is a correlation between high cost areas and higher uncertainty (variance) on the average delay cost that will be experienced by flights. 

\begin{figure}[h]
    \centering
    \includegraphics[width=0.5\linewidth]{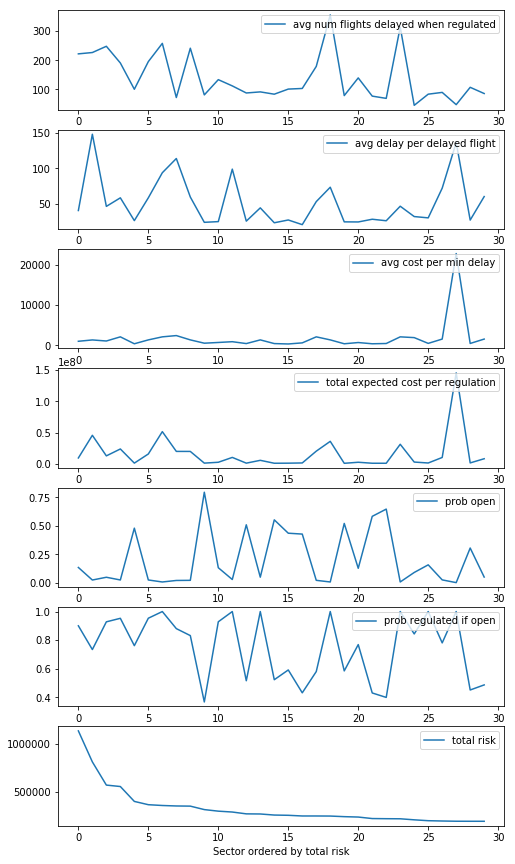}
    \caption{Parameters influencing the severity of sectors.}
    \label{fig:components}
\end{figure}

In the next step, the expected average costs ($\TC(s)$) per delayed flight are combined with the probability of being regulated. In other words, we apply Equation \ref{eq:all} to generate economic risk ($\TR(s)$) for each sector. To illustrate the impact of different components used in the calculation of economic risk, Figure~\ref{fig:components} presents their influence for the 30 sectors with highest total risk for the period 2016-2018. 

The subplots in Figure~\ref{fig:components} are ordered following the different parameters presented in the Equation \ref{eq:all}: average number of delayed flights when a regulation is issued in the sector, average delay per delayed flight and average cost per minute of delay. Note how these parameters change across sectors. Some sectors/regions might have flights operated with larger aircraft which on average have higher cost of delay or higher average delay per delayed flight. These parameters lead to the total expected cost per regulation (shown in the fourth subplot from the top). The expected costs when regulated does not lead directly to the total risks of the sector, as this is experienced when there is a regulation, therefore, we consider the probability of sector being open and probability of being regulated if the sector is open. Figure~\ref{fig:components} presents how these parameters combine to experience different levels of total risk. A given sector might have a larger expected cost per regulation but be open very seldom. Or, it might have a low probability for being regulated if open but being in use most of the time. 

The economic risk of sectors is then transformed into economic severity, using the relationship between the percentile of economic risk and economic severity category, as listed in Table~\ref{tab:economic_severity}. The economic severity indicator can be aggregated over different time horizons. 

\begin{figure*}[!t]
\centerline{
\subfloat[Quarter Q1]{\includegraphics[width=0.35\linewidth]{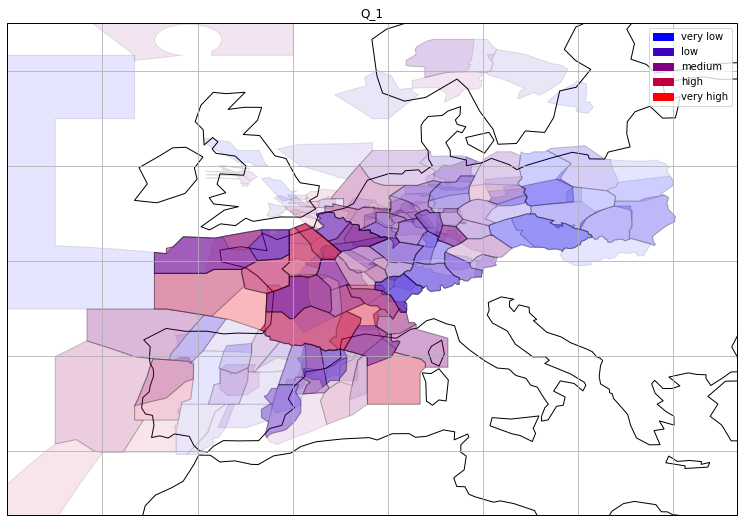}%
\label{subfig:severity_q1}}
\hfil
\subfloat[Quarter Q2]{\includegraphics[width=0.35\linewidth]{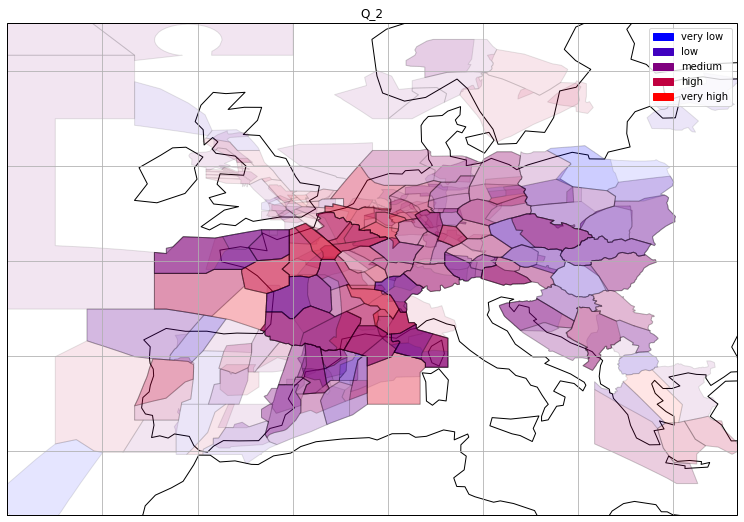}%
\label{subfig:severity_q2}}}
\centerline{
\subfloat[Quarter Q3]{\includegraphics[width=0.35\linewidth]{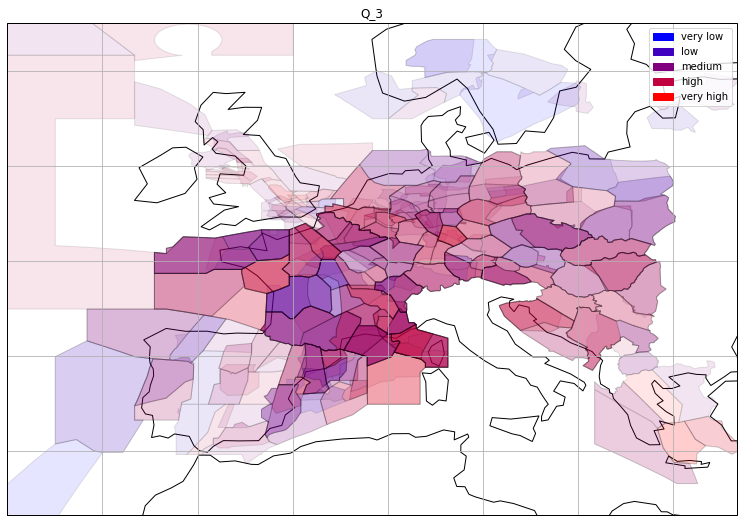}%
\label{subfig:severity_q3}}
\hfil
\subfloat[Quarter Q4]{\includegraphics[width=0.35\linewidth]{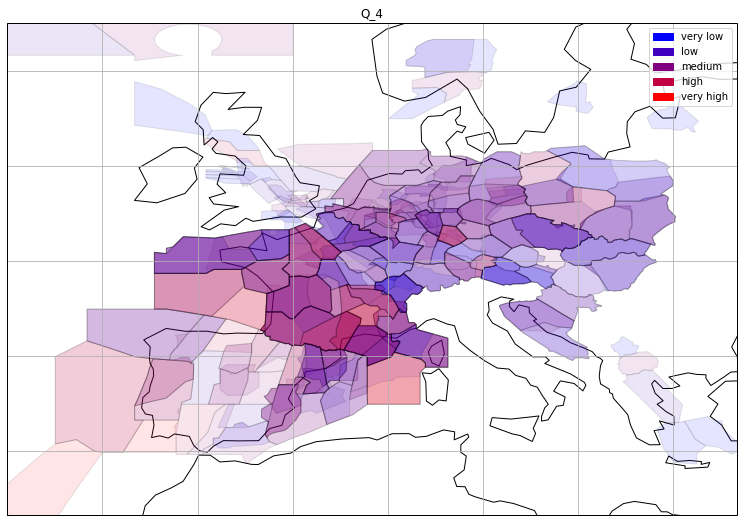}%
\label{subfig:severity_q4}}}
\caption{Severity of sectors, divided by quarters and aggregated across all years (2016-2018).}
\label{fig:severity_aggregated}
\end{figure*}

Figure~\ref{fig:severity_aggregated} presents the severity of different sectors which had at least one regulation in the period 2016-2018. The data has been aggregated per quarters for the whole 2016-2018 period: Q1 (January-March), Q2 (April-June), Q3 (July-September), Q4 (October-December). This particular aggregation gives an indication on how the situation changes over the seasons. Note how, as expected, Q2 and Q3 have the most high and very high severe sectors. This is aligned with a higher amount of traffic in summer season. We expect to see a correlation between severity and high demand and complexity, as more complex airspace with higher demand is more likely to be regulated. It is worth noticing how some areas in Europe might not be affected by regulations at different periods of the year (e.g., Balkan region in Q1). In the presented examples, we are computing the severity of different sectors by analysing the ATFM regulations issued on them. However, in some cases, ATFM regulations apply to other airspace elements, beyond sectors, such as Traffic Volumes (TV). These TVs might not have a unique relationship with sectors and, in those cases, they are not included in this severity analysis. For this reason, some areas, such as the Italian airspace seem not to have regulated regions in the figures. The framework presented in this paper could be extended to other airspace elements in the future.

\begin{figure*}[h]
\centerline{
\subfloat[2016 Q3]{\includegraphics[width=0.35\linewidth]{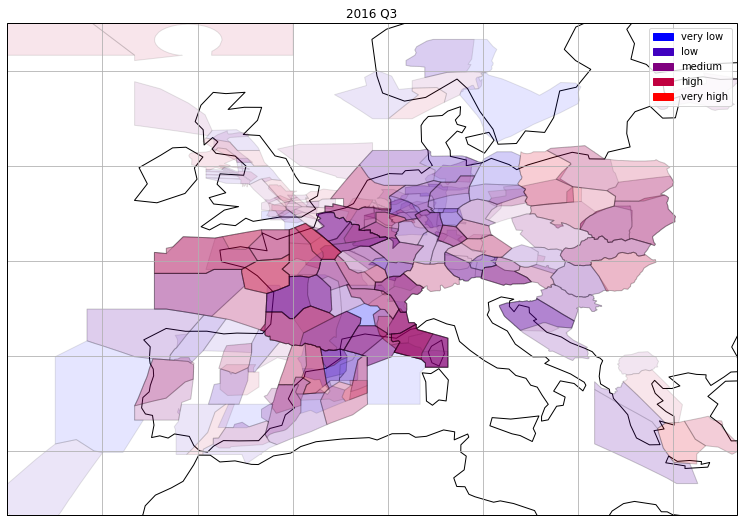}%
\label{subfig:severity_2016_q3}}
\hfil
\subfloat[2017 Q3]{\includegraphics[width=0.35\linewidth]{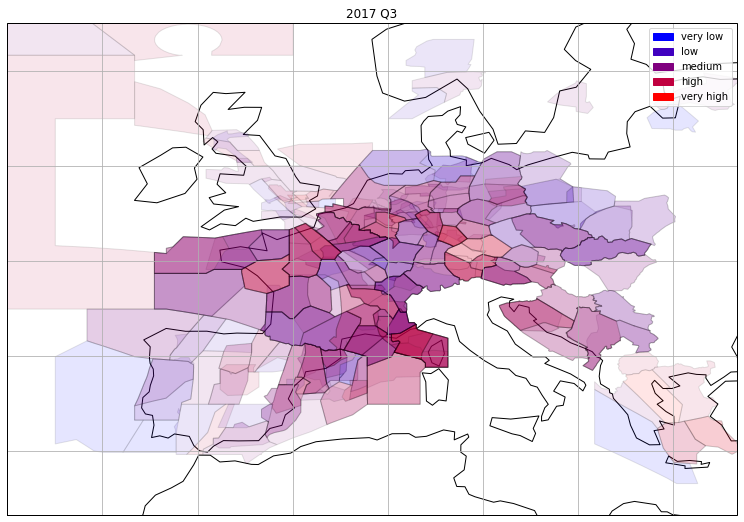}%
\label{subfig:severity_2017_q3}}}
\centerline{
\subfloat[2018 Q3]{\includegraphics[width=0.35\linewidth]{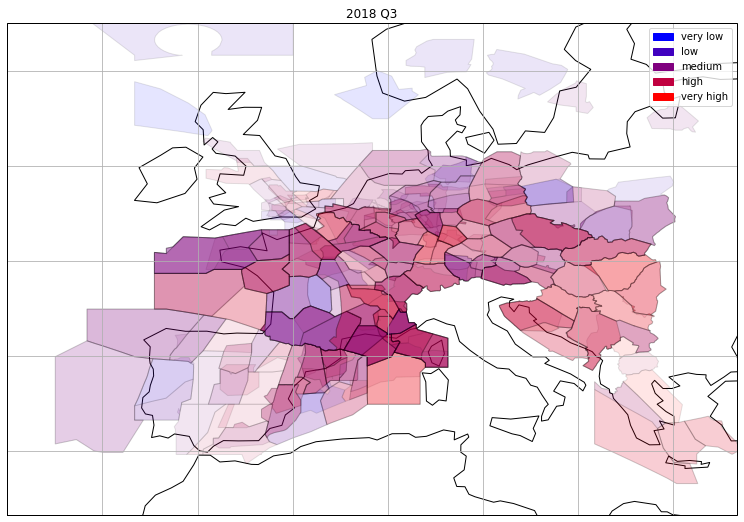}%
\label{subfig:severity_2018_q3}}}
\caption{Severity of sectors across 2016-2018 Q3}
\label{fig:severity_q3}
\end{figure*}

Figure~\ref{fig:severity_q3} presents the severity of the sectors in Europe for Q3 quarters across the 2016-2018 period. In this example, we aggregate the data to calculate economic risk and consequently the economic severity for each quarter in the period under analysis. The same criteria to classify the sectors into the different severity categories as in the previous case has been used. It is interesting to observe how the number of sectors in the higher categories of severity has increased over the 2016-2018 period. This is aligned with the trend of increasing traffic, number of regulations and consequent ATFM delay in Europe over the analysed period \citep{CODA_2018}.

\section{Discussion and conclusions}\label{sec:conclusions}

In current operations, there is a mismatch in time horizons between the capacity planning by ANSPs and the flight planning by airspace users, which often results in capacity-demand imbalances. These imbalances lead to ATFM regulations, delays, and consequently costs. The European ATM Master Plan~\citep{MasterPlan2020} envisions earlier sharing of information and collaborative planning, through the application of trajectory based operations and rolling network operations plan~\citep{NOP2019}. With the use of available data and machine learning techniques, we can envision different improvements in the information sharing aimed at collaborative decision making. 

In this paper we described the methodology behind the newly proposed metric of the \emph{economic severity} of airspace. The goal is to convey the information on the economic impact ATFM regulations have on different portions of airspace, to be used in the strategic flight planning. This metric aims at improving the information sharing between stakeholders and the decision making processes.

The economic severity is based on the \emph{economic risk} of an airspace element (e.g., sector), which is defined considering the cost a regulation in that airspace imposes on airspace users. The risk is computed using only historically available data: the delay generated by each airspace element based on the (historical) average number of flights delayed by regulations issued in that airspace, the average delay generated on those regulations, and the probability of having that specific airspace operational and, in that case regulated. The detailed estimation of the cost experienced by airspace users depends on various operational parameters. These parameters can be approximated by estimating the costs as a function of the \sMTOW of the aircraft and the delay assigned to each individual flight, capturing the non-linearity of delay and costs. However, these detailed data (delay issued per individual aircraft) might not be available strategically (as then data is usually reported at a higher level of aggregation). To overcome this limitation, in this paper, a relationship between average delay and average cost is developed. We demonstrated that a simple quadratic model ($\bar{c} = \alpha \bar{\delta} + \beta \bar{\delta}^2$, with $\alpha=23.1$ \euro{} per minute and $\beta=1.05$ \euro{} per minute square) is able to capture this relationship. We also concluded that including the information on the regulation reason does not increase the estimation capacity of this cost of delay model. The quantitative \emph{economic risk} can then be transformed into qualitative \emph{economic severity} by grouping the values of economic risk in economic severity categories.

The information the economic severity provides could help ANSPs and airspace users. ANSPs could identify regions posing higher economic risk and therefore require further operational improvements. The improvements can take different forms, such as  using different opening schemes at different times of day or even designing new sectorisations to better capture traffic demand complexities. To be able to vary opening schemes, ANSPs have to match them with the available controllers, thus impacting the controller rosters (within ANSP impact) and offered airspace capacity (external impact). Note that by focusing on cost and not on delay, we are able to provide an airspace user-centric view of the impact regulations have on the system, another important step in mutual understanding between the ANSPs, NM and airspace users. The temporal evolution of the system is captured by selecting different historical time-frames to compute the severity of the sectors. Airspace users could, in their turn, also benefit from this view of the network. Adjusting the time-frame used for the analysis can help them to identify regions which could pose a higher risk to their operations and to consider this strategically to minimise their impact (e.g., adjusting buffers on specific flight segments).

To be able to use the metric operationally, two things are deemed necessary: first, further development of the visualisation of the rolling network operations plan; second, consultation with the stakeholders to define the categorisation of the economic risk into economic severity. 

\begin{figure}[htbp]
    \centering
    \includegraphics[scale=0.6]{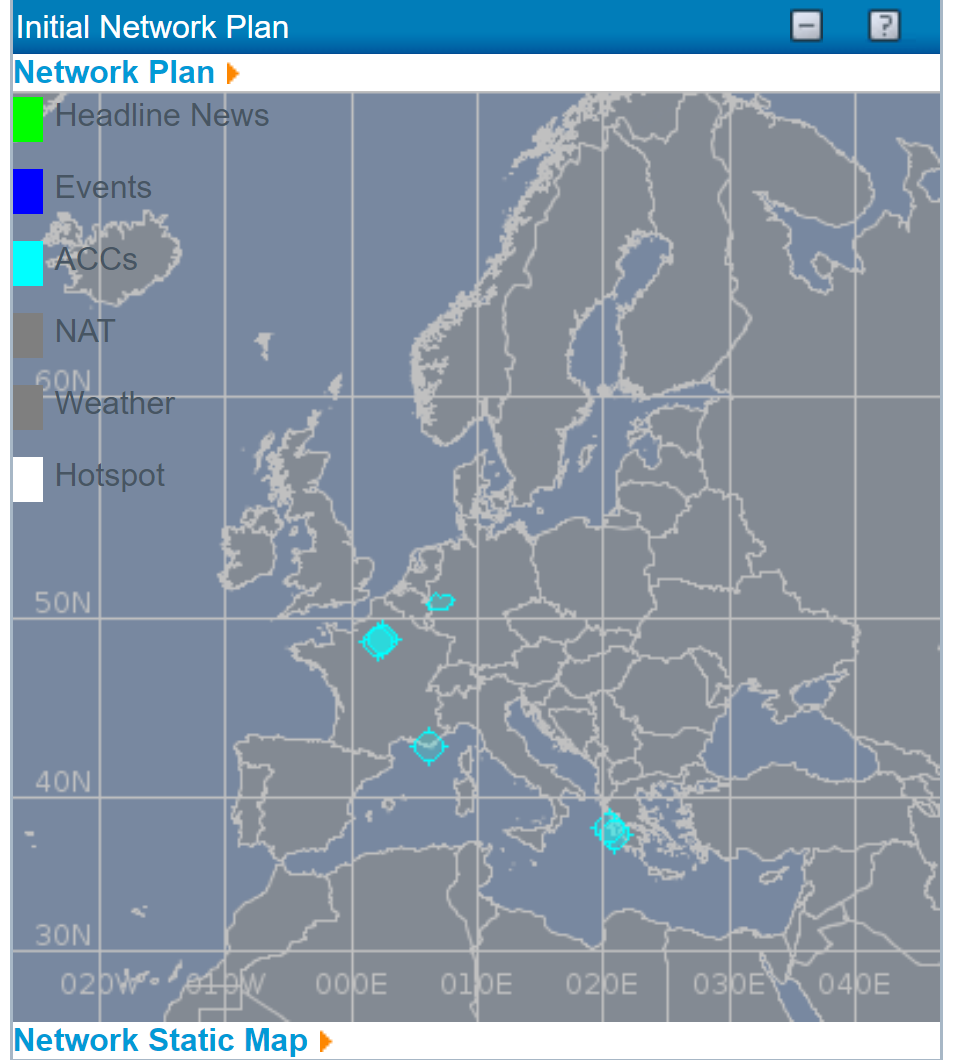}
    \caption{Screenshot of the current initial network plan available on the EUROCONTROL's public Network Operations Portal. }
    \label{fig:nop}
\end{figure}

As presented in Figures~\ref{fig:severity_aggregated} and \ref{fig:severity_q3} the economic severity can easily be used in visualisation tools, thus being easily shared among different stakeholders. Figure~\ref{fig:nop} shows the current version of the initial network plan at the EUROCONTROL's public Network Operations Portal, intended for the pre-tactical time frame. The current rolling seasonal network operations plan~\citep{ECTRL2021} is available in the form of a long document, listing planned interventions on the ATM infrastructure, forecast traffic, bottlenecks, mitigation actions and other concerns. In the future, we could imagine having a visual representation of the (strategic) rolling network  plan, overlaying the economic severity visualisation on such a map, similar to the one shown in Figure~\ref{fig:nop}.  

In this paper, the economic risk is classified using quantiles into five easily understood categories: very low, low, medium, high and very high. It might happen that the suggested categorisation is not the most suitable for the stakeholders. For example, the airspace users might be interested only in the very high risk areas. This characteristic of the economic severity metric should be further explored with the stakeholders.

We presented two types of data aggregation for metric calculation: one for different quarters across three years and another one on quarter by quarter basis. It is very probable that the quarter by quarter calculation is the most useful for the stakeholders, as it avoids conflating other influences (e.g., traffic increase or decrease, weather). In some cases, however, a monthly aggregation might be preferred. Feedback from stakeholders is required to select the most suitable time-frame.

To sum up, we defined the methodology for economic severity metric, the model for cost of delay to be included in the metric, and presented the results on the three years of ATFM regulation data. The ultimate goal behind this work is to use the knowledge obtained from data to better the strategic planning-decision processes. Future work would entail the stakeholder consultations to further fine-tune the details that would help make the metric usable in the stakeholder operations.

\section*{Acknowledgment}
This work has been performed as part of the ADAPT project which has received funding from the SESAR Joint Undertaking under grant agreement No 783264 under European Union’s Horizon 2020 research and innovation programme.

\section*{Glossary}

\textbf{Acronym} - \textbf{Definition}\\
AIRAC - Aeronautical Information Regulation And Control\\
ANSP -	air navigation service provider \\
ATC	- air traffic control \\
ATFCM - air traffic flow control management \\
ATFM -	air traffic flow management \\
ATM - air traffic management \\
C - ATC capacity regulation type \\
CASA - Computer Assisted Slot Allocation \\
DDR2 -	demand data repository 2 \\
ECDF - experimental cumulative distribution \\
G - aerodrome capacity regulation type \\
GBT - gradient boosting tree \\
I - industrial action regulation type \\
M - airspace management regulation type \\
MTOW - maximum take-off weight \\
NM - Network Manager \\
O - other regulation type \\
S - ATC staffing regulation type \\
TV - traffic volume \\ 
W - weather regulation type\\

\bibliographystyle{apalike}
\bibliography{regulation}

\begin{thebibliography}{}

\bibitem[Boli{\'c} et~al., 2017a]{Bolic2017a}
Boli{\'c}, T., Castelli, L., Corolli, L., and Rigonat, D. (2017a).
\newblock Reducing atfm delays through strategic flight planning.
\newblock {\em Transportation Research Part E}, 98:42--59.

\bibitem[Boli{\'c} et~al., 2017b]{bolic2017peak}
Boli{\'c}, T., Castelli, L., and Rigonat, D. (2017b).
\newblock Peak-load pricing for the european air traffic management system
  using modulation of en-route charges.
\newblock {\em European Journal of Transport and Infrastructure Research},
  17(1).

\bibitem[Cook and Tanner, 2015]{Cook2015}
Cook, A.~J. and Tanner, G. (2015).
\newblock European airline delay cost reference values.
\newblock Technical report, EUROCONTROL Performance Review Unit.

\bibitem[Delgado et~al., 2015]{controller2015}
Delgado, L., Cook, A., Crist\'{o}bal, S., and Plets, H. (2015).
\newblock {Controller Time and Delay Costs - a trade-off analysis}.
\newblock In {\em Proceedings of the Fifth SESAR Innovation Days (2015),
  EUROCONTROL}.

\bibitem[EUROCONTROL, 2017]{Eurocontrol2017}
EUROCONTROL (2017).
\newblock {Annual Network Operations Report 2016}.
\newblock Technical report, {EUROCONTROL}.

\bibitem[EUROCONTROL, 2018a]{CFMUmanual}
EUROCONTROL (2018a).
\newblock {ATFCM Users Manual}.
\newblock Technical report, {EUROCONTROL}.

\bibitem[EUROCONTROL, 2018b]{CODA_2018}
EUROCONTROL (2018b).
\newblock {CODA Digest 2018 -- All causes delay and canellations to Air
  Transport in Europe}.
\newblock Technical report, {EUROCONTROL}.

\bibitem[{EUROCONTROL}, 2019]{NOP2019}
{EUROCONTROL} (2019).
\newblock {European Network Operations Plan 2019-2024}.
\newblock Technical report, {EUROCONTROL}.

\bibitem[EUROCONTROL, 2019]{Eurocontrol2019}
EUROCONTROL (2019).
\newblock {Network Operations Report 2018}.
\newblock Technical report, {EUROCONTROL}.

\bibitem[EUROCONTROL, 2020]{Eurocontrol2020}
EUROCONTROL (2020).
\newblock {Network Operations Report 2019}.
\newblock Technical report, {EUROCONTROL}.

\bibitem[EUROCONTROL, 2021]{ECTRL2021}
EUROCONTROL (2021).
\newblock European network operationsplan 2021 rolling seasonal plan, edition
  10.
\newblock Technical report, EUROCONTROL.

\bibitem[{European Parliament and the Council}, 2013]{EU2004}
{European Parliament and the Council} (2013).
\newblock {Regulation (EC) No 261/2004 of the European Parliament and of the
  Council of 11 February 2004 establishing common rules on compensation and
  assistance to passengers in the event of denied boarding and of cancellation
  or long delay of flights}.
\newblock Technical report, {European Parliament and the Council}.

\bibitem[Friedman, 2001]{friedman_greedy}
Friedman, J.~H. (2001).
\newblock {Greedy function approximation: A gradient boosting machine}.
\newblock {\em {Annals of Statistics}}, 29(5):1189--1232.

\bibitem[Ivanov et~al., 2019]{ivanov2019}
Ivanov, N., Jovanović, R., Fichert, F., Strauss, A., Starita, S., Babić, O.,
  and Pavlović, G. (2019).
\newblock Coordinated capacity and demand management in a redesigned air
  traffic management value-chain.
\newblock {\em Journal of Air Transport Management}, 75:139 -- 152.

\bibitem[{Ivanov} et~al., 2017]{Ivanov17}
{Ivanov}, N., {Netjasov}, F., {Jovanovic}, R., S.{Starita}, and {Strauss}, A.
  (2017).
\newblock {Air Traffic Flow Management slot allocation to minimize propagated
  delay and improve airport slot adherence}.
\newblock {\em {Transportation Research Part A}}, 95:183--197.

\bibitem[Jovanovi{\'c} et~al., 2014]{Jovanovic2014}
Jovanovi{\'c}, R., To{\v{s}}i{\'c}, V., {\v{C}}angalovi{\'c}, M., and
  Stanojevi{\'c}, M. (2014).
\newblock Anticipatory modulation of air navigation charges to balance the use
  of airspace network capacities.
\newblock {\em Transportation Research Part A: Policy and Practice}, 61:84--99.

\bibitem[{Montlaur} and {Delgado}, 2020]{Delgado20}
{Montlaur}, A. and {Delgado}, L. (2020).
\newblock Flight and passenger efficiency-fairness trade-off for atfm delay
  assignment.
\newblock {\em Journal of Air Transport Management}, 83.

\bibitem[{Network Manager}, 2018]{NM2018}
{Network Manager} (2018).
\newblock {\em IFPS users manual}.
\newblock EUROCONTROL.

\bibitem[Nosedal et~al., 2014]{NOSEDAL2014171}
Nosedal, J., Piera, M.~A., Ruiz, S., and Nosedal, A. (2014).
\newblock An efficient algorithm for smoothing airspace congestion by
  fine-tuning take-off times.
\newblock {\em Transportation Research Part C: Emerging Technologies}, 44:171
  -- 184.

\bibitem[Nosedal et~al., 2015]{NOSEDAL201511}
Nosedal, J., Piera, M.~A., Solis, A.~O., and Ferrer, C. (2015).
\newblock An optimization model to fit airspace demand considering a
  spatio-temporal analysis of airspace capacity.
\newblock {\em Transportation Research Part C: Emerging Technologies}, 61:11 --
  28.

\bibitem[Schultz et~al., 2018]{aerospace5040109}
Schultz, M., Lorenz, S., Schmitz, R., and Delgado, L. (2018).
\newblock Weather impact on airport performance.
\newblock {\em Aerospace}, 5(4).

\bibitem[{SESAR Joint Undertaking}, 2020]{MasterPlan2020}
{SESAR Joint Undertaking} (2020).
\newblock {European ATM Master Plan - Executive view}.
\newblock Technical report, {SESAR Joint Undertaking}.

\bibitem[Starita et~al., 2020]{starita2020}
Starita, S., Strauss, A.~K., Fei, X., Jovanovi{\'c}, R., Ivanov, N.,
  Pavlovi{\'c}, G., and Fichert, F. (2020).
\newblock {Air Traffic Control Capacity Planning Under Demand and Capacity
  Provision Uncertainty}.
\newblock {\em Transportation Science}, 54(4):882--896.

\bibitem[Tang et~al., 2012]{TANG201289}
Tang, J., Alam, S., Lokan, C., and Abbass, H.~A. (2012).
\newblock A multi-objective approach for dynamic airspace sectorization using
  agent based and geometric models.
\newblock {\em Transportation Research Part C: Emerging Technologies}, 21(1):89
  -- 121.

\bibitem[Xu et~al., 2020a]{Xu2020}
Xu, Y., Dalmau, R., Melgosa, M., Montlaur, A., and Prats, X. (2020a).
\newblock A framework for collaborative air traffic flow management minimizing
  costs for airspace users: Enabling trajectory options and flexible
  pre-tactical delay management.
\newblock {\em Transportation Research Part B: Methodological}, 134:229--255.

\bibitem[Xu et~al., 2020b]{XU2020359}
Xu, Y., Prats, X., and Delahaye, D. (2020b).
\newblock Synchronised demand-capacity balancing in collaborative air traffic
  flow management.
\newblock {\em Transportation Research Part C: Emerging Technologies}, 114:359
  -- 376.

\end{thebibliography}

\end{document}